\documentclass[onecolumn,11pt,secnumarabic,amssymb, nobibnotes, aps, prl, superscriptaddress, citeautoscript, noeprint]{revtex4-2}
\usepackage[utf8]{inputenc}
\usepackage[colorlinks=true,linkcolor=blue,citecolor=blue,urlcolor=blue]{hyperref}
\usepackage{graphicx} % Required for inserting images
\DeclareUnicodeCharacter{00A0}{ }	
\usepackage[separate-uncertainty=true]{siunitx}
\usepackage[labelfont=bf]{caption}
\usepackage{titlesec}
\usepackage{setspace}
\usepackage[dvipsnames]{xcolor}
\usepackage[a4paper, total={7in, 7in}]{geometry}
\usepackage{url}
\usepackage{gensymb} % for \degree
\usepackage{amsmath}
\usepackage{ulem}

\setcitestyle{super}

\DeclareCaptionLabelSeparator{bar}{ \textbar{} }
\titleformat{\section}
  {\raggedright\bf}{}{0pt}{}
\titleformat{\subsection}
  {\raggedright\bf}{}{0pt}{}

\titlespacing*{\section}{0pt}{\baselineskip}{0pt}
\titlespacing*{\subsection}{0pt}{0pt}{0pt}

\newcommand{\comment}[1]{}

\begin{document}
\title{Correlations drive the attosecond response of strongly-correlated insulators}
%Attosecond probing of electronic correlations in a strongly-correlated insulator}
% Impact of strong correlations on the attosecond reponse of insulators
% Correlations drive the attosecond response of strongly-correlated insulators
%Dynamical electronic correlations drive the strong-field response of a correlated insulator}
%Attosecond to metastable dynamics in a strongly-correlated insulator}
%Dynamical electronic correlations drive the light-induced response of a strongly-correlated insulator
%Attosecond spectroscopy of a strongly-correlated insulator
%Attosecond-resolved dynamical electronic correlations in a strongly-correlated insulator}
\author{Romain Cazali}
\thanks{These two authors contributed equally}
\affiliation{Universit\'{e} Paris-Saclay, CEA, LIDYL, 91191 Gif-sur-Yvette, France}
\author{Amina Alic}
\thanks{These two authors contributed equally}
\affiliation{Sorbonne Universit\'{e}, CNRS, Laboratoire de Chimie Physique - Mati\`{e}re et Rayonnement, LCPMR, 75005 Paris, France}
\author{Matthieu Guer}
\affiliation{Universit\'{e} Paris-Saclay, CEA, LIDYL, 91191 Gif-sur-Yvette, France}
\author{Christopher J. Kaplan}
\affiliation{Department of Chemistry, University of California, Berkeley, 94720, USA}
\author{Fabien Lepetit}
\affiliation{Universit\'{e} Paris-Saclay, CEA, LIDYL, 91191 Gif-sur-Yvette, France}
\author{Olivier Tcherbakoff}
\affiliation{Universit\'{e} Paris-Saclay, CEA, LIDYL, 91191 Gif-sur-Yvette, France}
\author{St\'{e}phane Guizard}
\affiliation{Universit\'{e} Paris-Saclay, CEA, LIDYL, 91191 Gif-sur-Yvette, France}
\affiliation{CY Cergy Paris Universit\'e, CEA, LIDYL, 91191 Gif-sur-Yvette, France}

\author{Angel Rubio}
\affiliation{Max Planck Institute for the Structure and Dynamics of Matter, Luruper Chaussee 149, 22761 Hamburg, Germany}
\affiliation{Center for Free-Electron Laser Science CFEL, Deutsches Elektronen-Synchrotron DESY, Notkestra\ss e 85, 22607 Hamburg, Germany}
\affiliation{Center for Computational Quantum Physics (CCQ), The Flatiron Institute, 162 Fifth Avenue, New York NY 10010, USA}

\author{Nicolas Tancogne-Dejean}
\affiliation{Max Planck Institute for the Structure and Dynamics of Matter, Luruper Chaussee 149, 22761 Hamburg, Germany}
\affiliation{Center for Free-Electron Laser Science CFEL, Deutsches Elektronen-Synchrotron DESY, Notkestra\ss e 85, 22607 Hamburg, Germany}

\author{Gheorghe S. Chiuzb\u{a}ian}
\affiliation{Sorbonne Universit\'{e}, CNRS, Laboratoire de Chimie Physique - Mati\`{e}re et Rayonnement, LCPMR, 75005 Paris, France}
\author{Romain G\'{e}neaux}
\email{romain.geneaux@cea.fr}
\affiliation{Universit\'{e} Paris-Saclay, CEA, LIDYL, 91191 Gif-sur-Yvette, France}
\affiliation{CY Cergy Paris Universit\'e, CEA, LIDYL, 91191 Gif-sur-Yvette, France}

\begin{abstract}
% 200 mots max (actuel = 153)
Attosecond spectroscopy of materials has provided invaluable insight into light-driven coherent electron dynamics. However, attosecond spectroscopies have so far been focused on weakly-correlated materials. As a result, the behavior of strongly-correlated systems is largely unknown at sub- to few-femtosecond timescales, even though it is typically the realm at which electron-electron interactions operate. Here we conduct attosecond-resolved experiments on the correlated insulator nickel oxide, and compare its response to a common band insulator, revealing fundamentally different behaviors. The results, together with state-of-the art time-dependent \textit{ab initio} calculations, show that the correlated system response is governed by a laser-driven quench of electron correlations. The evolution of the on-site electronic interaction is measured here at its natural timescale, marking the first direct measurement of Hubbard U renormalization in NiO. It is found to take place within a few femtoseconds, after which structural changes slowly start to take place. The resulting picture sheds light on the entire light-induced response of a strongly-correlated system, from attosecond to long-lived effects.
\end{abstract}

\maketitle
\noindent Strongly correlated materials present peculiar behavior of both technological and fundamental interest \cite{Tokura2008, Paschen2021}, which stems from the strong repulsive electron-electron interaction, as well as strong coupling between charge, spin, orbital, and lattice degrees of freedom. Because of these highly intertwined order parameters, excitation with laser pulses offers striking control possibilities, such as light-induced phase transitions \cite{Cavalleri2001, delaTorre2021, Johnson2023} and superconductivity \cite{Cavalleri2018, Fava2024}, access to non-thermal states \cite{Stojchevska2014, Maklar2023} and the creation of Floquet-Bloch states \cite{Bloch2022}. The Hubbard $U$ parameter quantifies the effective on-site Coulomb repulsion between electrons, a key factor governing correlated electronic systems. In transition metal oxides, $U$ has values of several electron-volts, meaning that correlated electron dynamics are expected to take place on timescales of $\hbar/U \simeq 0.1$--$1$~fs \cite{Silva2018}. Therefore, strong correlations might facilitate the control of parameters and phases of materials at exceptionally high speeds, using light fields with optical cycles of a few femtoseconds.\\\\ 
Understanding non-equilibrium dynamics in correlated materials at ultrashort timescales is crucial for advancing technologies reliant on ultrafast electronic switching and phase control. Despite their transformative potential, the response of these systems under strong light fields remains poorly understood. Theoretically, standard first-principle calculations suffer from difficulties when adapting existing adiabatic pictures to highly out-of-equilibrium dynamics. Experimentally, accessing these timescales requires attosecond resolution. While state-of-the-art attosecond experiments can now explain the sub-cycle response of weakly-correlated semiconductors \cite{Schultze2014, Inzani2023} and band insulators,\cite{Schultze2013,Lucchini2016} most existing descriptions rely on single-particle pictures or simplified few-band models. For instance, the dynamical Franz-Keldysh effect \cite{Jauho1996, Lucchini2016, Volkov2023}, a key transient effect at the attosecond scale, is explained as a field-induced  modification of the single-particle band structure. This description is expected to break down when many-body processes dominate the optical response of the materials. Therefore, the response of strongly correlated solids to intense light fields on few- and sub-femtosecond timescales remains unknown. Do electronic correlations amplify the sub-optical-cycle dynamics under an external light pulse, or do they suppress the ultrafast response by enhancing electromagnetic field screening?\\ 
% note: maybe add something about attosecond photoemission delays & correlation in gases, e.g. Ossiander Nat Phys 2017 or Biswas arXiv:2111.14464?

In order to answer this question, we perform attosecond-resolved measurements on a band insulator and a correlated insulator, in exactly the same conditions. We choose two simple test-bed materials: magnesium oxide, a prototypical weakly-correlated metal oxide, and nickel oxide, which is historically the first system identified as a strongly-correlated insulator \cite{Mott1949}. Unlike some other correlated materials, \cite{Cavalleri2001} nickel oxide does not undergo any light-induced phase transition and thus offers the advantage of decoupled electronic and structural orders at short times.
Our results reveal completely different light-induced responses between the two oxides at the attosecond timescale: while MgO shows only transient reversible effects, NiO exhibits purely long-lived changes without visible subcycle dynamics. This shows that existing intuitive single-particle based pictures of insulators cannot be applied to nickel oxide, and fail to predict its response to intense light excitations. Instead, the response of NiO is explained by a light-driven quench of the on-site correlations, meaning that the intense laser field dynamically reduces the electron-electron interactions, as revealed by our \textit{ab initio} simulations based on real-time TDDFT complemented by an effective Hubbard~$U$\cite{Tancogne-Dejean2018,Tancogne-Dejean2020}. Following this interpretation, the measurement gives a direct access to the light-driven renormalization of the Hubbard $U$ on its true timescale, evidencing an ultrafast electronic response of only a few femtoseconds. Additionally, we follow the NiO response up to \qty{50}{\pico\second} and observe the relaxation corresponding to the formation of a structurally distorted non-thermal lattice state.
These measurements represent the first experimental observation of the birth of a non-thermal state driven by correlations, with the observation of the sub-femtosecond correlated electronic response, up to the long-lived picosecond dynamics of the metastable state induced by light.

\section*{Comparing correlated and band insulators}
\begin{figure}[htbp]
\includegraphics[width=15cm]{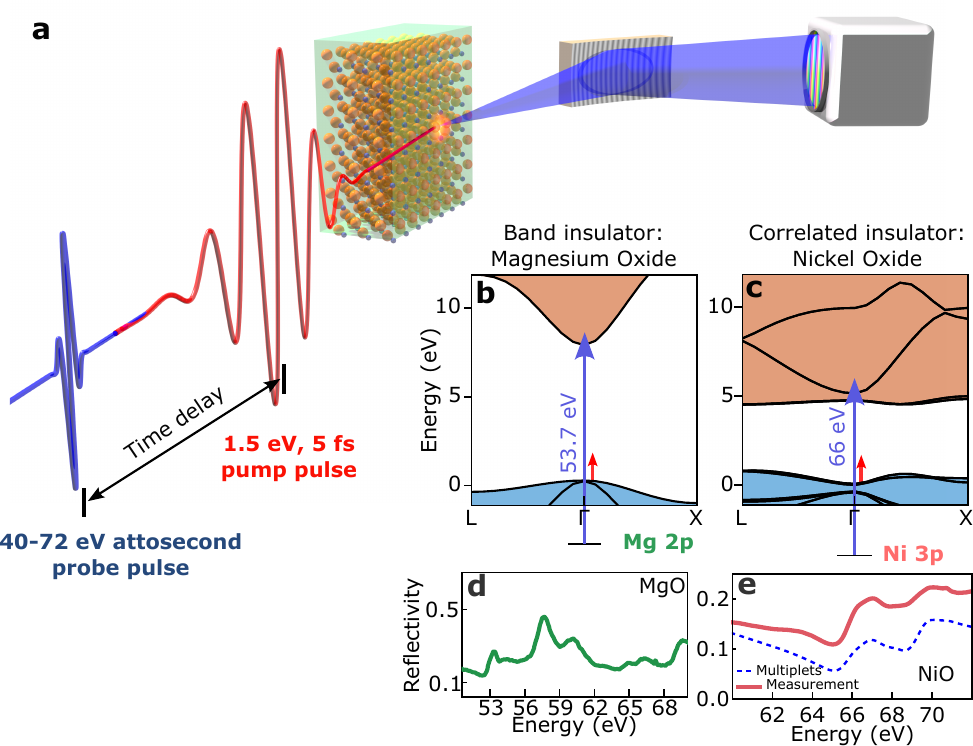}%V10 with multiplets, V9bis without
\centering
 \caption{\textbf{Principle of the experiment on band and correlated insulators.} \textbf{a,} Principle of the pump–probe scheme. \textbf{b, c,} Band structure obtained by LDA for MgO and LDA+U for NiO showing the XUV transition from the Mg $2p$ and Ni $3p$ core levels to their respective conduction bands. The parabolic shape of the MgO conduction band contrasts with the flat band of NiO, the latter being characteristic of strong correlations. \textbf{d, e, }
Measured static reflectivities of both samples, showing the Mg $L_{2,3}$ and the Ni $M_{2,3}$ edges. }%Les incertitudes sigma/sqrt(N) sont comprises dans l'épaisseur du trait...
\label{fig1}
\end{figure}

\noindent To explore the light-driven response of NiO and MgO, we perform pump-probe experiments according to the attosecond transient reflectivity scheme \cite{Kaplan2018}. Attosecond pulses with spectra spanning 40--72 eV impinge on the monocrystalline sample with an angle of 65\degree{ }from normal (see Fig \ref{fig1}a, Methods and SM for details on the experimental setup). Fig. \ref{fig1}d,e display the static reflectivity of the samples, showing either the Mg $L_{2,3}$ or the Ni $M_{2,3}$ edges (procedure detailed in SM). The sample is then excited by a p-polarized 5~fs laser pulse with an incident intensity of \qty{6}{\tera\watt\per\square\cm} and a central photon energy of 1.58 eV. The pump-probe delay is actively stabilized, guaranteeing a stability of 100~as over the course of the measurement. Fig. \ref{fig1}b,c displays the band structure of both systems around the $\Gamma$ point, illustrating that the bandgaps (7.8~eV for MgO\cite{roessler1967electronic} and 4.3~eV for NiO\cite{Hufner1994}) %remember to add in Methods that the LDA BS of MgO is shifted
are larger than the pump energy. It shows a key difference between the two insulators: the first NiO conduction band (which corresponds to the $3d$ $e_g$ orbitals) is much flatter than the one of MgO, corresponding to more localized electron density in real space and thus stronger correlations. Valence bands have strong O-$2p$ character in both materials, and are therefore not visible in the time-resolved data when probing from the Ni core level.

\begin{figure}[htbp]
\includegraphics[width=\textwidth]{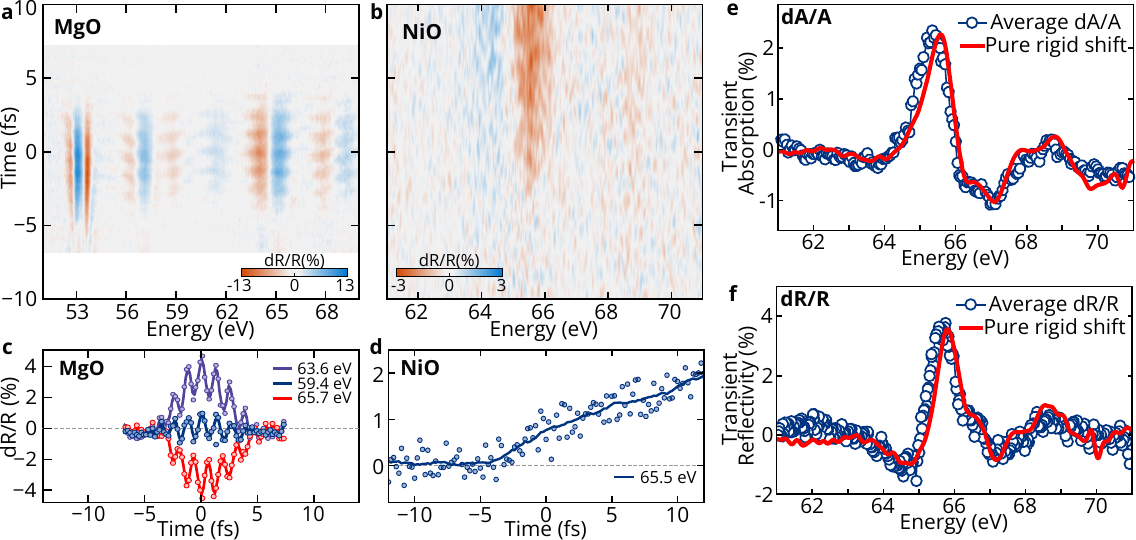}
\centering
\caption{\textbf{Attosecond transient reflectivity of MgO and NiO.} \textbf{a, b,}. Experimental transient reflectivity traces for MgO and NiO, as a function of pump-probe delay and XUV energy. Note the very different color scales for the two traces. \textbf{c, d,} Temporal slices of the transient reflectivty for Mgo and NiO at selected energies. \textbf{e, }The experimental transient absorption (open circles, obtained by Kramers-Kronig analysis) is very well reproduced by a pure redshift of the edge absorption (solid red line). Here the red shift is \qty{-51\pm2}{\milli\electronvolt}. \textbf{f, } The experimental transient reflectivity (open circles) is also very well reproduced by a pure redshift (solid red line) of the edge reflectivity. Here the red shift is also \qty{-51\pm2}{\milli\electronvolt}.}
\label{fig2}
\end{figure}

Starting with the more conventional case of the band insulator MgO, Figure \ref{fig2}a shows its attosecond transient reflectivity. The strong driving field induces large reflectivity change ($\approx 20$\%) localized around the temporal overlap, as previously reported in Ref~\cite{geneaux_attosecond_2020}. In addition, oscillations with 1.3~fs period are resolved across the entire spectrum (Fig.~\ref{fig2}c), corresponding to half the driving electric field period. This is typical of the Dynamical Franz-Keldysh Effect (DFKE), which has been identified in the band insulators diamond \cite{Lucchini2016} and $\text{SiO}_2$ \cite{Volkov2023}. Such oscillations have also been reproduced by more general semiclassical theories \cite{Cistaro2021}. The DFKE is a field-driven change of the absorption of a material near its bandgap, giving rise to shifts and satellite features that both oscillate in synchrony with the laser electric field. It is noteworthy that this effect is accurately described by perturbation theory applied to the static electroabsorption coefficient \cite{Jauho1996}, which is well-suited to weakly-correlated materials such as MgO. In addition, here the signal extends slightly more in the negative time direction, which is a signature of core excitation  with few-femtosecond lifetimes, as analyzed in details in $\text{SiO}_2$ \cite{Moulet2017} and $\text{MgF}_2$ \cite{Lucchini2020}. While a detailed study of these oscillations is not the topic of this paper, the measurement effectively exemplifies the behavior of a bandgap insulator, and provides an important reference measurement of the instrument temporal resolution.

Keeping identical experimental conditions -- pump fluence, duration, and probing scheme -- Figure \ref{fig2}b displays the same quantity for NiO. Figure \ref{fig2}d shows a temporal lineout of the main observed feature. The response of NiO has several significative differences compared to MgO: (i) the pump-induced reflectivity change is much weaker ($\approx 4$\%), despite its smaller band gap, (ii) the material response rises more slowly, and persists after excitation and (iii) no field-driven oscillations are resolved when the pump pulse is active. These observations already clearly suggests that the two systems behave fundamentally differently upon photo-excitation. The differences with MgO cannot simply be explained by their different bandgaps, as the DFKE was observed in materials with much smaller gaps (e.g. 1.42~eV for GaAs \cite{Schlaepfer2018}). We note that at reduced pump intensity, the NiO response decreased in magnitude, but kept the same characteristics and spectral shape (see SM). In order to explain the response of NiO, we perform a detailed analysis of the spectral changes of the Ni $M_{2,3}$~edge. Because the measurement is performed in reflection geometry, we apply a Kramers-Kronig treatment of the data in order to extract the absorption coefficient. The absolute reflectivity of the sample (Fig.~\ref{fig1}c) is padded with literature data to cover the visible to x-ray range~\cite{geneaux_attosecond_2020}, before applying a variational Kramers-Kronig algorithm~\cite{Kuzmenko2005}. The obtained absorption coefficient matches well with synchrotron measurements \cite{Chiuzbuaian2005}.
Figure \ref{fig2}e compares the  obtained transient absorption at positive delays with a pure redshift of \qty{-51}{\milli\eV}, displaying a remarkable match. At all time delays, the transient absorption trace is extremely well reproduced by a pure rigid shift of the absorption edge. The absence of more complex features, such as broadening or the appearance of new band-like signals, suggest that the signal does not stem from carrier photoinjection \cite{Zurch2017a} or heating \cite{Chang2021}. Because the spectral change is so simple, even the transient reflectivity signal is well described by a pure redshift of the same amount, as shown in Figure \ref{fig2}f. This allows us to conclude that the pump-induced response at the M-edge can be described purely as a redshift. 

\section*{Ultrafast renormalization of Hubbard U }
This observation evokes the possibility that the main effect at play is a light-induced shift of the NiO conduction band (CB). In strongly correlated systems, this has been shown to be a consequence of a laser-induced reduction of the Hubbard $U$, thereby shifting the upper Hubbard band (UHB, which is the CB in NiO) down in energy. Although $U$ renormalization in NiO was predicted theoretically \cite{Tancogne-Dejean2018, Tancogne-Dejean2020}, free-electron-laser experiments have not managed to observe it so far \cite{Wang2022, Granas2022, Lojewski2024}. We attribute this to the longer excitation pulses used in those studies, which prevented reaching the same excitation regime and lacked the temporal resolution required to distinguish the effect. This is apparent in our experiment, in which the transient signal can only be simply quantified as a pure redshift during the first \qty{40}{\femto\s}. After that, other processes come into play and the interpretation becomes less obvious. On the other hand, the clear transient renormalization of $U$ was observed experimentally in a transition metal dichalcogenide \cite{Beaulieu2021} and in a cuprate superconductor \cite{Baykusheva2022}, but so far without the possibility to directly follow its temporal evolution. In order to explore this possible scenario, we perform advanced theoretical simulations. Modeling the experiment accurately is quite challenging: it requires describing out-of-equilibrium electronic motion in the presence of strong correlations and non-linear optical excitation. In addition, it requires computing the time-dependent XUV reflectivity with sufficient time-resolution. We resort to time-dependent density functional theory plus Hubbard U (TDDFT +U), including all necessary core levels, coupled to macroscopic Maxwell equations. This approach provides a good balance between computational efficiency and accuracy in describing real systems~\cite{PhysRevB.96.245133}. We use the same pump pulse duration and fluence as in the experiment, making the calculation entirely \textit{ab initio}. Figures~\ref{fig3}a,b compare the transient reflectivity signal measured experimentally and obtained theoretically with the same color scale. The agreement is striking, reproducing the absence of oscillations in the transient reflectivity and the exact spectral change corresponding to the redshift. The only discrepancy is that the theory predicts a faster rise time of the signal compared to the experiment, which will be discussed further below. To evidence the role of the dynamic evolution of the Hubbard $U$ parameter, we repeat the calculations while keeping $U$ frozen to its ground-state value -- effectively removing these dynamical correlations and keeping other physical effects identical. Importantly, the frozen $U$ calculation still includes all population transfer and local-field effects. Figure \ref{fig3}c displays the resulting transient reflectivity trace, with the signal almost entirely disappearing. Comparing the spectral changes more precisely (Figure \ref{fig3}d) we demonstrate that the experimental redshift is almost entirely due to the renormalization of $U$. We thus conclude that other effects, typically relevant for weakly-correlated materials, only marginally affect the out-of-equilibrium response of NiO. 

Given the ability of the TDDFT+U simulations to faithfully reproduce the experimental transient, it is instructive to complement them with a simpler local model based on multiplet calculations. These calculations, widely used to interpret transition-metal core-level spectra \cite{Laan2006}, rely on model Hamiltonians that incorporate atomic and ligand-field effects. Although not \textit{ab initio} and formally limited to an adiabatic description of the electronic structure, they offer two advantages that are particularly relevant here: (i) they allow us to identify which local parameters control the observed spectral shift, and (ii) they enable rapid exploration of parameter space, providing intuition for how the transient changes in electronic structure relate to the microscopic mechanisms.
To ensure consistency with experiment, we first determine a set of multiplet parameters that reproduce the static $M_{2,3}$ reflectivity (Fig.~\ref{fig1}e and SM). In the temporal regime of interest, we assume that the crystal field remains effectively unchanged, and vary only parameters directly associated with correlation and charge transfer. Following the TDDFT+U results, the Hubbard $U$ is transiently reduced by $\Delta U =\SI{-100}{\milli\electronvolt}$ and accompanied with a corresponding renormalization of the charge-transfer energy by $\Delta U/2$. Within these constraints, we find that a rigid shift of the $M_{2,3}$ edge comparable to the measured one cannot be achieved by modifying $U$ alone: an adjustment of the Ni–O hybridization strength is also required (Fig.~\ref{fig3}e). Quantitatively, a \SI{3.4}{\percent} reduction of the hybridization reproduces the experimental shift for $\Delta U =\SI{-100}{\milli\electronvolt}$. More generally, we find a linear relation between the required $U$ renormalization and hybridization change: for $\Delta U$ between 200 and \qty{0}{\milli\electronvolt}, a hybridization variation of $(0.0085\Delta U + 4.2)\si{\percent}$ (with $\Delta U$ given in \si{\milli\electronvolt}), yields a satisfactory match (see SM\cite{SuppMat}). This analysis does not aim to provide a fully predictive description of the dynamics but rather to identify which local parameters must evolve to account for the measured spectral shift. Within the multiplet approach, the experimentally observed redshift can be interpreted using a minimal local picture involving a simultaneous renormalization of $U$ and the Ni–O hybridization. This establishes a simple framework for thinking about the transient electronic structure of NiO and lays the groundwork for future approaches of correlated materials.

\section*{Few-femtosecond screening dynamics}
\begin{figure}[htbp]
\includegraphics[width=15cm]{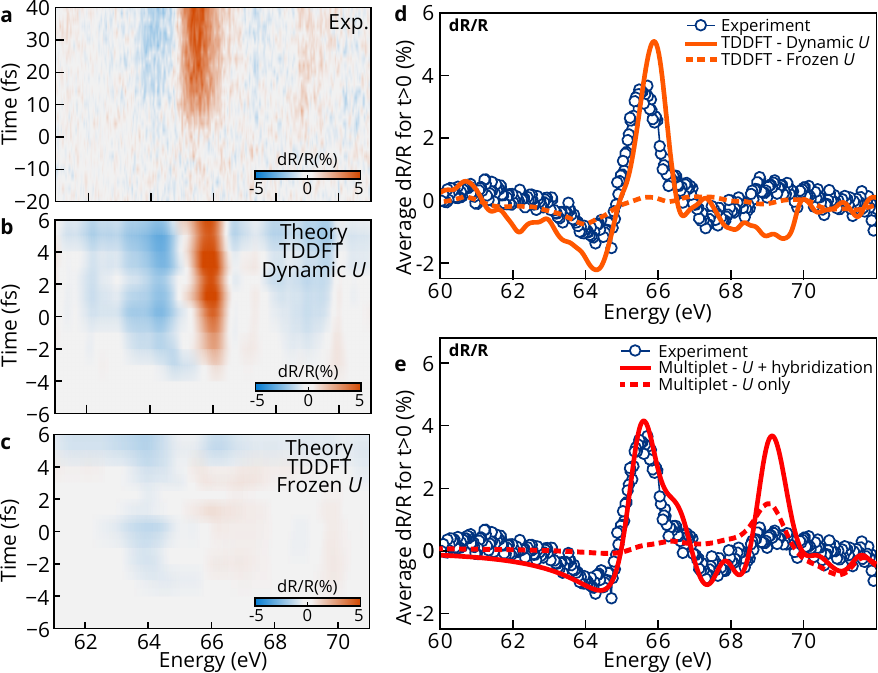}
\centering
\caption{\textbf{Importance of dynamical electronic correlations}. \textbf{a,} Measured transient reflectivity up to \qty{40}{\femto\second}, showing the full build-up of the signal. \textbf{b,} Calculated transient reflectivity by TDDFT+U calculations when allowing dynamical electronic correlations. Aside from the different timescale (note the vertical axis), the agreement is excellent both in shape and in magnitude. \textbf{c,} Same calculation as in \textbf{b}, but now considering a frozen $U$. The color scale is kept the same as in \textbf{a} and \textbf{b} for comparison. \textbf{d,} Averaged transient reflectivity for positive time delays, in the case of the experiment (solid blue line), TDDFT with dynamic U (solid orange line), and TDDFT with frozen U (dashed red line). \textbf{e,} Same as \textbf{d} but using multiplet calculations, with either a reduction of U only (using the TDDFT value), or with a reduction of hybridization as well (to match the experimental redshift).}
\label{fig3}
\end{figure}

\begin{figure}[htbp]
\includegraphics[width=\textwidth]{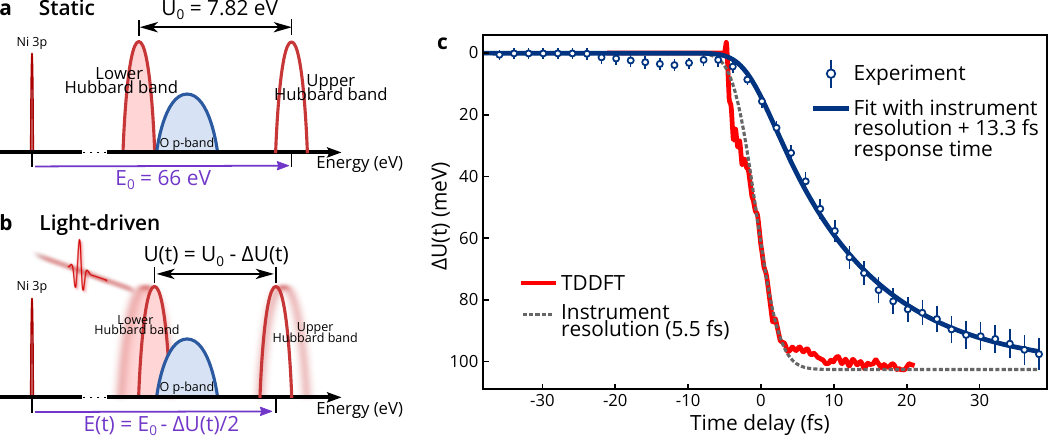}
\centering
\caption{\textbf{Timescale of the Hubbard U renormalization}. \textbf{a,} Schematic band diagram of the system: at equilibrium, the measured signal corresponds to transition from the Ni $3p$ core-level to the empty upper Hubbard band, which is separated by $U_0$ from the lower Hubbard band. \textbf{b,} Upon photoexcitation, $U$ becomes time-dependent and the energy of the XUV transition is shifted by $\Delta U(t)/2$. \textbf{c,} Dynamic evolution of $U(t)$, obtained experimentally (circles) and theoretically (solid red line). The error bars correspond to $\pm1$ standard deviations. The solid blue line is a fit of the measurement taking into account the experimental time resolution and an inherent system response time - found here to be \qty{13.3\pm1.0}{\femto\s}. The grey dashed line shows the time resolution of the experiment.}
\label{fig4}
\end{figure}

Going back to the experimental results with the $U$ renormalization in mind, we can directly interpret the data to measure the temporal change in the Hubbard parameter, $\Delta U(t)$. As illustrated in Figure \ref{fig4}a,b, $\Delta U(t)$ brings the upper (resp. lower) Hubbard band down (resp. up) in energy by $\Delta U(t)/2$. Therefore, the energy of the $\text{Ni}\,{3p} \rightarrow \text{UHB}$ transition, which is the probe step of the experiment, will be reduced by $\Delta U(t)/2$. The aforementioned redshift thus corresponds to half the change in $U$. Figure \ref{fig4}c presents the experimentally obtained $\Delta U(t)$, using a least-square fit at each time delay to get the shift value, and compares it with the one calculated from our TDDFT+U. The agreement in their final magnitude (\qty{-100}{\milli\electronvolt}) is excellent, confirming again the accuracy of the simulation. The discrepancy, as seen before in Figure \ref{fig3}, lies in an underestimated response time of the system. Experimentally, this response time is comprised of the instrument response function and of the intrinsic property of the material. The instrumental time resolution is cross-checked using both the transient reflectivity of MgO and the transient absorption of helium, a common reference system in attosecond spectroscopy \cite{Ott2014}. In both cases, the resolution is found to be \qty{5.5}{\femto\s}, which is the expected order of magnitude given the duration of the pump pulse. As can be seen Figure \ref{fig4}c, the measured response of NiO is significantly slower than the time resolution. Fitting the signal with a functional including a non-instantaneous system response and accounting for the time resolution \cite{deRoulet2024}{} (see SM), yields a intrinsic response time of \qty{13.3\pm1.0}{\femto\s} for NiO. In order to understand the origin of this non-instantaneous response, we point out that the TDDFT calculations predict an essentially immediate material response, limited only by the duration of the laser pulse (see Fig. \ref{fig4}c). Therefore, the origin of the \qty{13.3}{\femto\s} response time lies in a degree of freedom that is not included in our TDDFT calculations. There are two main candidate scenarios: first, the dynamics could involve bosonic degrees of freedom such as phonons or magnons, which are not included in the calculation. In NiO, the fastest excitations are zone-center LO and TO phonons, which have periods of 56 and \qty{84}{\femto\s}, respectively \cite{Kant2009}, which is significantly longer than the measured response time. Besides, we observe distinct signatures of structural dynamics at longer time delays, as will be further explained below. Therefore, the second and more likely scenario is the purely electronic delayed response, as electrons do not react instantly to the light-induced change in screening. The functional used in our TDDFT+U calculations is not designed to reproduce such response, because it is equivalent to an adiabatic approximation, meaning that it forces the electronic system to react instantly to the local screening. However, delayed electronic responses have been for instance observed by THz spectroscopy in GaAs \cite{huber2001}. One thus expects a timescale to be associated to the change of dielectric screening, and for uncorrelated electrons, this can be estimated from the plasma frequency \cite{huber2001}. State-of-the-art extended dynamical mean-field theory (EDMFT) has also demonstrated that this is also the case for correlated systems \cite{Murakami2023}. For instance, Golez et al.\ \cite{Golez2015} showed that in a single-band $U-V$ Hubbard model subjected to an electric field pulse, the system's response is delayed, occurring on the timescale of electron hopping. While these theoretical approaches are not directly comparable to our measurements, they strongly support that the non-instantaneous response time is a key feature of the solid. How correlations affect this timescale remains a challenging question and will require theoretical developments.

%Going back to figure \ref{fig2}.a we see that a change in the Hubbard parameter is directly probed by the XUV probe in the form of a shift in the energy of the transition from the 3p core orbital to the upper Hubbard band. Since we observe a redshift we can deduce that the Hubbard parameter changes by $\Delta U = 2 \Delta E$.  This means that we have a direct experimental access the the change of $U$ in real time.
The consequences of this few-femtosecond excitation were then investigated in measurements up to \qty{50}{\pico\second}. After \qty{40}{\femto\second}, the transient reflectivity can no longer be described as a pure redshift, meaning that the Hubbard $U$ can no longer be tracked (see raw data and fitting in the Supplemental Material \cite{SuppMat}). Instead, we observe a blue shift of the differential reflectivity at the Ni edge with a timescale of $\tau_1$ = \qty{280 \pm 60}{\femto \s}, together with the appearance of new features at 40-45 eV with a timescale of $\tau_2$ = \qty{260 \pm 30}{\femto \s}. Similar spectral features have been observed in $\text{Fe}_2\text{O}_3$~\cite{Carneiro2017} and NiO~\cite{Biswas2018, Biswas2018b} and attributed to charge-transfer excitation from oxygen to the metal center which triggers a shift of the oxygen atoms away from the metal and results in the formation of a small polaron. In our excitation conditions, the charge transfer gap can be reached by three-photon absorption: extracting the photoinduced electron-hole pair density from the TDDFT+U simulation yields $N \sim 2.7 \times 10^{20}$\,cm$^{-3}$, corresponding to a population of 0.6\% of electron-hole pairs per unit cell. This is comparable to the density of excited carriers produced by above-gap excitation (4.66 eV) of NiO at $\sim 400~\mu\text{J.cm}^{-2}$ \cite{rossi2025}. 

The $\tau_1$ timescale is also comparable to that reported by a later time-resolved electron diffraction experiment in NiO\cite{Windsor2021} using sub-gap excitation, which found that the rhombohedral lattice distortion along [111] — present in the ground state due to exchange striction — is reduced within \qty{310 \pm 80}{\femto\second}. The similarity of electronic and structural timescales confirms the assignment to charge transfer followed by polaron formation. Afterwards, the amplitude of the signal decays with $\tau_3$ = \qty{6\pm0.7}{\pico\s}, and remains constant within our sensitivity for up to \qty{50}{\pico\second}. The final metastable state is therefore found to be very long-lived, showing that here, strong correlations do not lead to fast thermalization. Such thermalization bottlenecks were already identified in charge-transfer and Mott insulators \cite{Kollath2007, Eckstein2009, rossi2025}.

%\section*{Conclusion}
In summary, we have demonstrated that the correlated insulator NiO exhibits a fundamentally different attosecond response to an intense laser field, compared to the band insulator MgO. We do not observe any field-driven $2\omega$ signal in either experiment or simulations, despite speculation of its existence in earlier works \cite{Baykusheva2022, Granas2022}. This brings forth important questions for attosecond science, such as the feasibility of lightwave control at optical frequencies \cite{Borsch2023} in strongly-correlated systems, in contrast to standard insulators where this was demonstrated in numerous studies\cite{Hanus2021, Ossiander2022}. To further strengthen this observation, 
we also performed CEP-dependent TDDFT simulations (see SM \cite{SuppMat}) and find that the system response does not substantially change with CEP. This demonstrates that the Hubbard $U$ renormalization -- the preeminent system response -- is driven by the laser intensity and not by its electric field. 

More generally, this work provides the first time-resolved observation of light-driven interaction quenching at its natural timescale. We evidence an inherent response time of a few femtoseconds for the process, showing that the $U$ renormalization is extremely fast, but not quite instantaneous. This timescale is a new observable linked to the inherent response of the electron system, and represents the speed at which the electron-electron interaction, or screening, can be manipulated by a laser pulse. The ability to dynamically control the Hubbard $U$ on femtosecond timescales could be foundational for accessing and harnessing new nonequilibrium material states. While our simulations reproduce the magnitude of the response very well, calculating its timescale is so far out of reach. This will thus serve as a challenging benchmark for future \textit{ab initio} theories. Finally, this work provides valuable insight for the interpretation of ultrafast phemonena in strongly-correlated materials, such as high-harmonic generation \cite{Silva2018, Tancogne-Dejean2018, Shao2022, Uchida2022} or photo-induced phase transitions \cite{Jager2017, Johnson2023}, in which the role of a dynamically-changing $U$ remains to be further explored.

\clearpage
\section{Methods}
\subsection*{TDDFT+U simulations}
\noindent All the calculations presented here were performed for bulk NiO, which is a type-II antiferromagnetic material below its N\'eel temperature ($T_N=\SI{523}{\kelvin}$\cite{cracknell_space_1969}). We neglected the small rhombohedral distortions and considered NiO in its cubic rock-salt structure, which does not affect the result of calculated optical spectra. Calculations were performed including spin-orbit coupling using fully norm-conserving pseudo-potentials for Ni and O, generated thanks to the Ape code~\cite{oliveira2008generating}, treating explicitly 3$p$, 3$d$, and 4$s$ orbitals as valence orbitals for Ni, and 2$p$ orbitals for O. We employed a lattice parameter of 4.1704~\AA\, a real-space spacing of $\Delta r=0.3$ Bohr, and a $16\times16\times8$ $\mathbf{k}$-point grid to sample the Brillouin zone.
The driving field is taken along the [100] crystallographic direction in all the calculations. We consider a laser pulse of 5.2 fs duration (FWHM), matching the experimental duration, with a sin-square envelope for the vector potential. The experimental carrier wavelength $\lambda=760$\,nm was employed, corresponding to a carrier photon energy of 1.63\,eV. In all calculations, we set the carrier envelope phase (CEP) to zero. We checked, see SI, that the CEP as no impact on the light-induced of change of the Hubbard $U$.
The time-dependent wavefunctions, electric current, and $U_{\mathrm{eff}}$ are computed by propagating generalized Kohn-Sham equations within real-time TDDFT+U, as provided by the Octopus package~\cite{10.1063/1.5142502}.
We employed the LDA functional~\cite{PhysRevB.23.5048} for describing the semilocal DFT part, and we computed the effective $U_{\mathrm{eff}}=U-J$ for the O $2p$ ($U^{2p}_{\mathrm{eff}}$) and Ni $3d$ orbitals ($U^{3d}_{\mathrm{eff}}$), using localized atomic orbitals from the corresponding pseudopotentials~\cite{PhysRevB.96.245133}.
All calculations are propagated for 13.8 fs after the kick, to avoid spurious numerical effects. A Lorentzian broadening of $\eta=0.3$ eV is obtained by applying a mask to the time-dependent signal before taking the Fourier transform, to mimic the experimental broadening. \\
In order to allow for a quantitative comparison with the experiment, the choice of the intensity in the simulation is quite important, and hence requires to treat the coupling to Maxwell equations at a macroscopic level.
The experiment is performed at an intensity of $I_{\mathrm{ext}}\sim6.0\,$TW.cm$^{-2}$ in vacuum. From this intensity, we obtain the strength of the electric field in vacuum. Due to the experimental geometry and the incidence at the Brewster angle, the strength of the transmitted electric field is obtained from the vacuum one by multiplying it by $1/n$, where $n=2.33$ is the refractive index of NiO at 760\,nm. This determines the strength of the electric displacement $\mathbf{D}$ in matter. The total electric field acting on the electrons results from the addition of this external field $\mathbf{D}$ with the induced field $\mathbf{E}_{\rm ind}$. This induced field is obtained as in Ref.~\cite{Tancogne-Dejean2020} by coupling the generalized Kohn-Sham equations to the macroscopic Maxwell equations, allowing to capturing the effect of the macroscopic induced transverse electric field. For each time delay, we propagate the generalized time-dependent Kohn-Sham equations of a time-dependent density functional theory plus Hubbard $U$ (TDDFT+U) framework~\cite{PhysRevB.96.245133} coupled to the macroscopic Maxwell equation and we later ``kick'' the system at the given time-delay to obtain the optical spectra (see Ref.~\cite{10.1063/1.5142502}). The macroscopic induced vector potential $\mathbf{A}_{\mathrm{ind}}(t)$ is computed from the total electronic current $\mathbf{j}(t)$ (atomic units are used throughout this work):
\begin{equation}
 \frac{\partial^2}{\partial t^2}\mathbf{A}_{\mathrm{ind}}(t) = \frac{4\pi c}{V}\mathbf{j}(t),
 \label{eq:maxwell}
\end{equation}
where $V$ is the volume of the cell. In order to obtain the vector potential induced by the kick, we subtract from the $\mathbf{A}_{\mathrm{ind}}(t)$ a reference calculation obtained in presence of the driving pump laser field but no kick. The dielectric function is then computed from the knowledge of the kick itself and vector potentials induced by the kick.

The time-dependent generalized Kohn-Sham equation within the adiabatic approximation reads \footnote{The nonlocal part of the pseudopotential is omitted for conciseness}
\begin{eqnarray}
 i\frac{\partial}{\partial t}|\psi_{n,\mathbf{k}}(t)\rangle = \Big[\frac{(\hat{\mathbf{p}}-\mathbf{A}_{\mathrm{tot}}(t)/c)}{2} + \hat{v}_{\mathrm{ext}} + \hat{v}_{\mathrm{H}}[n(\mathbf{r},t)]\nonumber\\
 + \hat{v}_{\mathrm{xc}}[n(\mathbf{r},t)] + \hat{V}_{U}[n(\mathbf{r},t),\{n_{mm'}\}]\Big]|\psi_{n,\mathbf{k}}(t)\rangle,
\end{eqnarray}
where  $|\psi_{n,\mathbf{k}} \rangle$ is a Pauli-spinor Bloch state with a band index $n$, at the point $\mathbf{k}$ in the Brillouin zone, $\hat{v}_{\mathrm{ext}}$ is the ionic potential, $\mathbf{A}_{\mathrm{tot}}(t)$ is the total vector potential containing the induced one of Eq.~\ref{eq:maxwell},  $\hat{v}_{\mathrm{H}}$ is the Hartree potential, $\hat{v}_{\mathrm{xc}}$ is the exchange-correlation potential. $\hat{V}_{U}$ is the non-local operator for DFT+U that depends also on the occupation matrix of the localized subspace $\{n_{mm'}\}$, see Ref.~\cite{PhysRevB.96.245133} for more details, including the definition of $U$, $J$, and $\hat{V}_{U}$.

\subsection{Multiplet calculations}

Our simulations rely on the multiplet ligand field simulations carried out with Quanty Refs.~\cite{haverkort2012multiplet, quanty_documentation}. For the calculation of the M$_{2,3}$ transitions in the unperturbed case, we adopt the main parameters previously listed for NiO: Hubbard $U = \SI{7.3}{\electronvolt}$, crystal field parameter in octahedral symmetry $10Dq=\SI{0.56}{\electronvolt}$, charge-transfer energy $E_{CT} = \SI{4.7}{\electronvolt}$, hybridization of Ni~$3d$ states with O~$2p$ states $V_{eg} = \SI{2.06}{\electronvolt}$, $V_{t2g} = \SI{1.21}{\electronvolt}$ and \SI{120}{\milli\electronvolt} magnetic exchange interaction. Moreover, while conserving the ab initio calculated values for the $3p$ and $3d$ spin-obit coupling,~\cite{van1991m2, haverkort2012multiplet} the reduction factors applied to the atomic Slater integrals were adapted for the best description of the measured XAS spectrum\cite{Chiuzbuaian2005}. The latter issued the following reduction factors of $F^2$, $G^1$ and $G^3$ as \SI{0.70}{\percent}, \SI{0.66}{\percent} and \SI{0.90}{\percent}, respectively. To describe the out-of-equilibrium states, we propose that the variation of the on-site Hubbard repulsion and the charge-transfer energy are linked through the linear relationship $\Delta U = 2 \Delta E_{CT}$, in line with our \mbox{TDDFT +U} results. Furthermore, the variation of the hybridization strengths was carried out at a constant $V_{eg}/V_{t2g}$ ratio. This simplification is supported by the hypothesis that the local symmetry is conserved during the first dozen \si{\femto\second}.

\vspace{\baselineskip}
\subsection*{Experimental setup}
\noindent The NiO sample is a commercially available \SI[parse-numbers=false]{5\times5\times1}{\cubic\milli\metre} single crystal delivered by MaTecK GmbH (Germany). The measurement is based on a Ti:Sapphire amplifier delivering \qty{24}{\femto\s}, \qty{2}{\milli\joule} pulses at a repetition rate of \qty{1}{\kilo\hertz}. The pulses are spectrally broadened using a stretched hollow-core fiber with a diameter of \qty{400}{\micro\m} and a length of \qty{1.4}{\m} (few-cycle Inc.), filled with a helium gradient ranging from \qty{0}{\bar} to \qty{3.5}{bar}. The pulse is then compressed using chirped mirrors (PC70, Ultrafast Innovations), yielding a pulse duration of \qty{5}{\femto\second} FWHM from a homemade dispersion scan setup. The central wavelength is \qty{760}{\nano\meter} which corresponds to an optical cycle duration of \qty{2.5}{\femto\second}. A detailed scheme of the experiment can be found in the supplementary material \cite{SuppMat}. 80\% of the beam is focused in an argon gas cell to generate a continuous XUV spectrum spanning from \qty{40}{\electronvolt} to \qty{72.5}{\electronvolt} serving as the probe beam. An aluminum dot filter filters out the driving infrared laser and lets a ring of infrared light that is latter used for active stabilization of the delay. The probe is then focused using a toroidal mirror in a 2f-2f configuration. The remaining 20\%   serves as the pump beam. It is recombined co-linearly with the XUV beam using a doubly drilled mirror. The pump and probe are then focused on the sample, with a measured pump intensity of \qty{6}{\tera\watt\per\square\cm}. This intensity did not lead to sample damage, which was verified by making sure that the XUV reflectivity was not decreasing over time, and by visible inspection of the sample surface. Damage was observed for intensities $\geq$\qty{10}{\tera\watt\per\square\cm}. The infrared light is then filtered out, and the XUV spectrum is recorded by a spectrometer made of a grating and a XUV CCD camera (greateyes GmbH). Delay stabilization is achieved by locking onto spatial fringes made from the unused pump and probe infrared beams. We use an interferometric filter centered at \qty{790}{\nano\meter} with a width of \qty{10}{\nano\meter}  to increase the duration of the pulse significantly, which extend the range of measurement of the delay. The final delay stability is \qty{100}{\atto\second} over several hours. The CEP of the laser is not stabilized. However, this does not prevent the measurement of sub-cycle features as demonstrated by our MgO measurements and previous experiments. All data were acquired thanks to the open source python-based software PyMoDAQ \cite{weber2021pymodaq}, and analyzed using the airPCR algorithm with optimal parameters of 28 iterations and 35 principal components \cite{Facciala2021}.

\section{Data Availability}
The data presented in the manuscript are available from the authors upon request.

\section{Acknowledgements}
We would like to thank Iris Crassee for helpful discussions on RefFIT and Siarhei Dziarzhytski for valuable discussions and for providing the sample. This work was supported by the European Union (ERC, Spinfield, Project No. 101041074 and Horizon 2020 Programme No. EU-H2020-LASERLAB-EUROPE-654148), the French Agence Nationale pour la Recherche (under grants TOCYDYS, ANR-19-CE30-0015-01 and HELIMAG, ANR-21-CE30-0037) and the Investissements d’Avenir program of LabEx PALM (ANR-10-LABX-0039-PALM). AA acknowledges the financial support from the Doctoral School ED 388 Chimie-Physique et Chimie Analytique de Paris Centre.

\section{Author contributions}
R.C., A.A., M.G., S.G., S.G.C. and R.G. performed the experiments. S.G.C. and R.G. initiated the study. R.C., A.A., C.J.K. and R.G. analyzed the experimental data. F.L. and O.T. operated the laser system. A.R. and N.T.D. performed the TDDFT calculations and first proposed the physical interpretation. A.A. and S.G.C. performed the multiplet calculations. R.C., N.T.D. and R.G. wrote the first version of the manuscript, to which all authors contributed.

% perhaps interesting ref https://arxiv.org/pdf/2302.11391
%\bibliographystyle{aapmrev4-2}
\bibliography{references}

\end{document}

\clearpage
\clearpage %needed for two-page reference section
\setcounter{page}{1}
\renewcommand{\thetable}{S\arabic{table}}  
\setcounter{table}{0}
\renewcommand{\thefigure}{S\arabic{figure}}
\setcounter{figure}{0}
\renewcommand{\thesection}{S\arabic{section}}
\setcounter{section}{0}
\renewcommand{\theequation}{S\arabic{equation}}
\setcounter{equation}{0}
\onecolumngrid

\title{Supplementary information for: Correlations drive the attosecond response of strongly-correlated insulators}
\author{Romain Cazali}
\thanks{These two authors contributed equally}
\affiliation{Universit\'{e} Paris-Saclay, CEA, LIDYL, 91191 Gif-sur-Yvette, France}
\author{Amina Alic}
\thanks{These two authors contributed equally}
\affiliation{Sorbonne Universit\'{e}, CNRS, Laboratoire de Chimie Physique - Mati\`{e}re et Rayonnement, LCPMR, 75005 Paris, France}
\author{Matthieu Guer}
\affiliation{Universit\'{e} Paris-Saclay, CEA, LIDYL, 91191 Gif-sur-Yvette, France}
\author{Christopher J. Kaplan}
\affiliation{Department of Chemistry, University of California, Berkeley, 94720, USA}
\author{Fabien Lepetit}
\affiliation{Universit\'{e} Paris-Saclay, CEA, LIDYL, 91191 Gif-sur-Yvette, France}
\author{Olivier Tcherbakoff}
\affiliation{Universit\'{e} Paris-Saclay, CEA, LIDYL, 91191 Gif-sur-Yvette, France}
\author{St\'{e}phane Guizard}
\affiliation{Universit\'{e} Paris-Saclay, CEA, LIDYL, 91191 Gif-sur-Yvette, France}
\affiliation{CY Cergy Paris Universit\'e, CEA, LIDYL, 91191 Gif-sur-Yvette, France}

\author{Angel Rubio}
\affiliation{Max Planck Institute for the Structure and Dynamics of Matter, Luruper Chaussee 149, 22761 Hamburg, Germany}
\affiliation{Center for Free-Electron Laser Science CFEL, Deutsches Elektronen-Synchrotron DESY, Notkestra\ss e 85, 22607 Hamburg, Germany}
\affiliation{Center for Computational Quantum Physics (CCQ), The Flatiron Institute, 162 Fifth Avenue, New York NY 10010, USA}

\author{Nicolas Tancogne-Dejean}
\affiliation{Max Planck Institute for the Structure and Dynamics of Matter, Luruper Chaussee 149, 22761 Hamburg, Germany}
\affiliation{Center for Free-Electron Laser Science CFEL, Deutsches Elektronen-Synchrotron DESY, Notkestra\ss e 85, 22607 Hamburg, Germany}

\author{Gheorghe S. Chiuzb\u{a}ian}
\affiliation{Sorbonne Universit\'{e}, CNRS, Laboratoire de Chimie Physique - Mati\`{e}re et Rayonnement, LCPMR, 75005 Paris, France}
\author{Romain G\'{e}neaux}
\email{romain.geneaux@cea.fr}
\affiliation{Universit\'{e} Paris-Saclay, CEA, LIDYL, 91191 Gif-sur-Yvette, France}
\affiliation{CY Cergy Paris Universit\'e, CEA, LIDYL, 91191 Gif-sur-Yvette, France}

\maketitle
\renewcommand{\baselinestretch}{0.75}\normalsize
\tableofcontents
\renewcommand{\baselinestretch}{1.0}\normalsize

\newpage
\section{Experimental setup}
\begin{figure}[htbp]
\includegraphics[width=15cm]{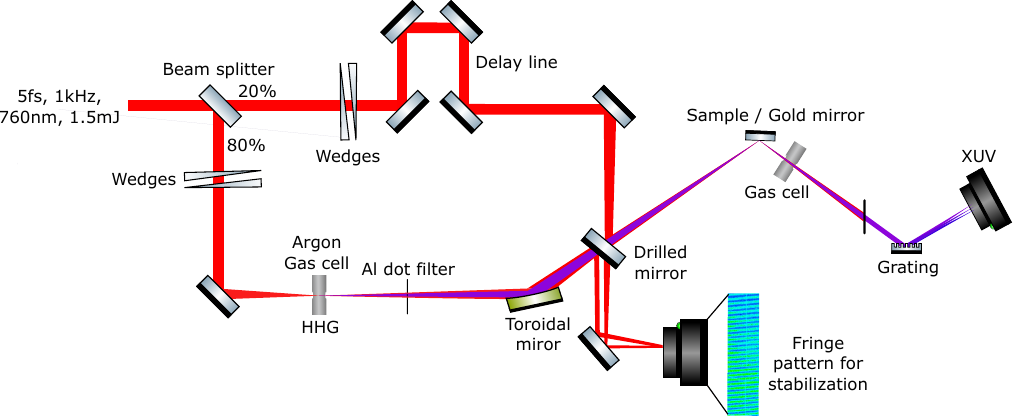}
\centering
\caption{\textbf{Experimental setup for attosecond transient reflectivity spectroscopy.} }. 
\label{fig:S1}
\end{figure}

A complete scheme of the experiment is represented on Figure \ref{fig:S1}. 

\subsection*{Energy calibration}
The energy axis was calibrated by inserting a gas cell in the XUV beam path after reflection from the sample. For calibration, we used the helium $2s^1 np^1$ %$3s 3p^6 np$  doubly excited state? I wrote Ar instead of He before...
series between \qty{60}{\eV} to \qty{66}{\eV}, and the neon $2s^1 2p^6 np$ series ranging from \qty{45}{\eV} to \qty{48}{\eV}. In addition, we used the aluminum absorption edge at \qty{72.5}{\eV} due to the aluminium filter. From these measurements, the quadratic relationship between wavelength and pixel position on the camera was determined.\\ 

\subsection*{Static reflectivity}
\begin{figure}[htbp]
\includegraphics[width=8cm]{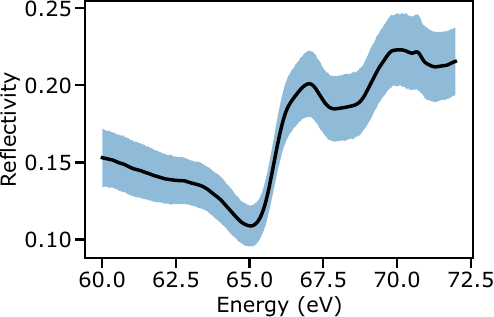}
\centering
\caption{\textbf{Absolute reflectivity measurement of NiO.} The blue shade represents the uncertainty at each energy obtained from the standard deviation $\sigma$ over 200 measurements. Note that the \qty{70.5}{\eV} feature comes from a defect on the camera, and is not a characteristic of NiO.}
\label{fig:S2}
\end{figure}

The static reflectivity of NiO was measured using both the reflected spectra of NiO ($R_{NiO}$) and of a gold mirror ($R_{Au}$). The energy axis of both the sample and the gold mirror was calibrated independently. To obtain the reflectivity of NiO relative to gold, we first calculated the ratio $R_{NiO}/R_{Au}$. Using literature values for the reflectivity of gold \cite{henke1993x}, we obtained the absolute reflectivity of NiO. To minimize errors due to the fluctuations of the harmonic source, we repeated the measurement 200 times. The standard error is given by $\frac{\sigma}{\sqrt{200}}$ where $\sigma$ is the standard deviation. The results are shown on Figure \ref{fig:S2}. The static reflectivity shows two main features at \qty{67}{\eV} and \qty{70}{\eV}. A small feature at \qty{70.5}{\eV} comes from a  dead pixel on the camera and is  not a characteristic of NiO.\\

\subsection*{Pump intensity}
\rc{The intensity of the pump beam was evaluated by measuring both its duration and the energy per unit of area at focus. We used an homemade in-situ dispersion scan to measure the pulse duration at the sample position, which routinely gave a full duration at half maxima of \qty{5}{\femto\s}. The energy per unit of area was determined by measuring the total energy in the incident pulse using a power meter, and by imaging the focus on a camera at the sample position. This allows calibrating the energy per unit area. The XUV focus is estimated to be \qty{60}{\micro\m} FWHM, thus we average the intensity over this area to obtain the average intensity probed during experiments. The non-normal incidence angle of the beams is taken into account with a projection of the pump beam profile on the sample, resulting in an intensity of \qty{6}{\tera\watt\per\square\cm}.  }

\section{Kramers-Kronig analysis}
\label{sec:KK}
Since XUV light is absorbed very efficiently by most materials, measuring the absorption at these wavelengths requires the use of ultra-thin samples, typically on the order of \qty{60}{\nano\m}. Obtaining  monocrystalline samples of this thickness is very challenging. Therefore, we used a bulk NiO in a reflectivity scheme instead of a direct absorption measurement. While absorption has a clear physical interpretation due to its direct relation with the density of state of the material, reflectivity depends on both the real and imaginary parts of the refractive index. The Kramers-Kronig relations link these components: they relate $\kappa$, the imaginary part of the refractive index, to $n$, the real part. Equivalently, they relate $\log(|r|)$, the absolute value of the reflectivity coefficient, and $\theta$, its phase. The computation of those relations requires the knowledge of the reflectivity over an infinite span of wavelength, which is not experimentally feasible. Thus, it is common to pad experimental data with data from the literature. Here we used two approaches,  the first is based on the direct computation of the Kramers-Kronig equation, and the second is based on a variational algorithm~\cite{Kuzmenko2005}.

The direct computation is based on the integration of the Kramers-Kronig relations for reflectivity $R$ and its phase $\theta$, following the procedure from Ref \cite{roessler1965kramers}. The equation for $\theta$ reads

\begin{equation}
    \theta = \int_0^{\infty} d\omega' \, \ln\left|\frac{\omega' + \omega}{\omega' - \omega}\right| \frac{d}{d\omega} \left[\ln\left(R(\omega)^{\frac{1}{2}}\right)\right].
\end{equation}

To compute the integral, the reflectivity was extended with the available data from 30~eV to 5000~eV. The reflectivity could have been extended up to 30000~eV, however no changes are observed by adding data above 5000~eV. Below 30~eV, reflectivity data was unavailable. Only absorption data exists, which cannot be used in the direct integral method. This is expected to change slightly the value of the retrieved index.

On the other hand, the variational method allows any input, absorption, reflectivity or the real part of the refractive index, allowing more literature data to be used. However this method leads to other difficulties, such as the right choice of model. In order to avoid discontinuities while doing the variational method, the static reflectivity curve was shifted so that the point at 60~eV matches literature data \cite{henke1993x}. In addition, this method has allowed us to include data below 30~eV as real or imaginary part of the refractive index.

\begin{figure}[htbp]
\includegraphics[width=17cm]{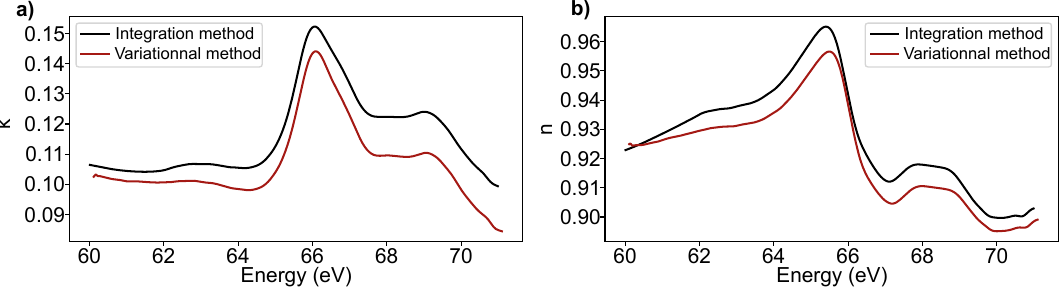}
\centering
\caption{Retrieved refractive index calculated from the static reflectivity using the direct computation (black curve) of the Kramers-Kronig integral and a variational algorithm (red curve). \textbf{a)} shows the imaginary index and \textbf{b)} shows the real index.}
\label{fig:S3}
\end{figure}
Retrieved $\kappa$  for both methods are shown on Figure \ref{fig:S3}. The two curves are very similar in shape, especially for the main feature around \qty{66}{\eV}, although they present some differences. The main difference between the two methods is a $20\%$ factor most likely due to the lack of data below 30~eV for the direct integration and the offset introduced for the variational method. This shows the limitations in retrieving the index quantitatively from a single reflectivity spectrum, which could only be avoided by more complex methods such as multi-angle measurements \cite{Kaplan2019b}. We conclude that both KK methods are consistent enough for retrieving the transient absorption, since the offset does not impact the pump-induced signal. The method is applied to the dynamical reflectivity at each time step, yielding the dynamical absorption, from which the transient absorption $dA/A$ is obtained.

\section{Redshift analysis} \label{redshift_analysis}
At each time-delay, two measurements are performed: one with the pump laser ($R_{on}$) and one without ($R_{off}$) allowing to calculate the transient reflectivity $\frac{dR}{R}= \frac{R_{on}-R_{off}}{R_{off}}$. The measurements were repeated between 10 and 50 times for each delay. The main source of noise is due to fluctuations of the harmonic source. This noise being very correlated on the entire spectra, an edge-referencing technique \cite{Geneaux2021} is systematically applied to significantly improve the quality of the measured signal.  Kramers-Kronig analysis gives the transient absorption $\frac{dA}{A}= \frac{A_{on}-A_{off}}{A_{off}}$. A pure red shift signal is simulated for all delays as $\frac{A_{static}(E+\Delta E)-A_{static}(E)}{A_{static}(E)}$ where $A_{static}(E)$ is the static absorption and $\Delta E$ is the shift in energy. To account for the background residual noise, an offset parameter is introduced to mitigate its effect. The resulting curve matches very well the measurement, proving that a pure red shift correctly explains the experiment. A pure redshift implies that both the real and the imaginary part of the refractive index undergo the exact same shift. This means that the same analysis with $\frac{dR}{R}$ should give the same result.  The results for both $\frac{dA}{A}$ and $\frac{dR}{R}$  are plotted in the main text on Figure 2. Although the two curves differ slightly in shape, the retrieved red shift are identical for both yielding a retrieved value of \qty{51\pm2}{\milli\eV}. Thus, in this particular case,  it is not necessary to perform a full Kramers-Kronig analysis in order to extract the red shift. If the signal was more complex or had multiple effects to disentangle, the absorption would be needed to avoid interpretation issues. Future work in this regime can directly use the reflectivity to retrieve the redshift and the Hubbard parameter change.

%\begin{figure}[htbp]
%\includegraphics[width=.8\textwidth]{figs/S4.pdf}
%\centering
%\caption{Pure redshift of $-51$meV for \textbf{a,} the transient absorption and \textbf{b,} the transient reflectivity. Although the curves slightly differ in shape, the associated redshift is identical.}
%\label{fig:S4}
%\end{figure}

\section{Fitting of the non-instantaneous electronic response}
In order to fit the evolution of $\Delta U(t)$ in Figure 4.c, we follow the procedure presented in de Roulet and coworkers \cite{deRoulet2024}. A simple error function is not able to describe a non-instantaneous system response, as it forces a symmetric behavior for negative and positive time delays. Instead, we consider an impulse response function of the form:
\begin{equation}
h(t, t_0, \tau) = \begin{cases}
0 &\text{for $t<t_0$}\\
1-\exp\left(-\frac{t-t_0}{\tau}\right) &\text{for $t\geq t_0$}
\end{cases}
\end{equation}
Where $t_0$ is the time corresponding to temporal overlap between pump and probe pulses, and $\tau$ is the response time of the system. The response function is then convolved with a Gaussian function with standard deviation $\sigma$:
\begin{equation}
f(t, t_0, \tau, \sigma) =\frac{1}{\sqrt{2\pi}\sigma} \int_{-\infty}^\infty \exp\left(-\frac{x^2}{2\sigma^2}\right) h(t-x, t_0, \tau) dx
\end{equation}
We now compute the integral to obtain a more practical expression, avoiding the convolution and the integration. The calculation is based on the following properties of the $\mathrm{erf}$ and $\mathrm{erfc}$ functions
\begin{align}
    \int_{-\infty}^{t} e^{-x^2} dx &=  \frac{\sqrt{\pi}}{2}
     (1+\operatorname{erf}(t))\\
    &= \frac{\sqrt{\pi}}{2} \operatorname{erfc}(-t)
\end{align}
where we used the definition $\mathrm{erfc}(t)=1-\mathrm{erf}(t)$ and the property $\mathrm{erfc}(t) = 2 - \mathrm{erfc}(-t)$. The function $f$ is decomposed in two integrals $f=\frac{I_1+I_2}{\sqrt{2\pi}\sigma}$ defined as 
\begin{equation}
    I_1 = \int_{-\infty}^{t-t_0}
    \exp\left(-\frac{x^2}{2\sigma^2}\right) 
     dx
\end{equation}
and 
\begin{equation}
    I_2 = -\int_{-\infty}^{t-t_0}
    \exp\left(-\frac{x^2}{2\sigma^2}\right) 
    \exp\left(-\frac{t - x - t_0}{\tau}\right)
     dx.
\end{equation}
The integral $I_1$ can be evaluated directly.  $I_2$ can also be computed directly after completing the square in the exponent 
\begin{equation}
    -\frac{x^2}{2\sigma^2}+\frac{x}{\tau} -\frac{t-t_0}{\tau} = -\left(\frac{x}{\sqrt{2}\sigma}
    -\frac{\sigma}{\sqrt{2}\tau}\right)^2+\frac{\sigma^2}{2\tau^2}-\frac{t-t_0}{\tau}.
\end{equation}
Combining the resulting expressions gives the following formula
\begin{equation}
    f(t) = \frac{1}{2} \left[   \text{erfc}\left(-\frac{t - t_0}{\sigma\sqrt{2}}\right)- \exp\left(\frac{\sigma^2}{2\tau^2}-\frac{t-t_0}{\tau}\right) \text{erfc}\left(-\frac{t - t_0}{\sigma\sqrt{2}} + \frac{\sigma}{\sqrt{2}\tau} \right) \right]\label{Analyticfitfunc}
\end{equation}
which is convenient for numerical calculation as it avoids both convolution and integration.
An alternative expression for this function,  as presented in Ref \cite{deRoulet2024} is given by
\begin{equation}
f(t, t_0, \tau, \sigma) =  \frac{1}{2\tau} \int_{- \infty}^t\exp\left[-\frac{(x-t_0)}{\tau}+\frac{\sigma^2}{2\tau^2}\right]  \times \mathrm{erfc}
\left[-\frac{(x-t_0)}{\sqrt{2}\sigma} + \frac{\sigma}{\sqrt{2}\tau}\right]dx\ldotp \label{fitfunc}\end{equation}
Our analytical formula can be recovered from this expression by applying integration by part and using previously mentioned properties of the error function. The first term $\exp \left[-\frac{(x-t_0)}{\tau}+\frac{\sigma^2}{2\tau^2}\right]$is integrated while the second term $\mathrm{erfc}
\left[-\frac{(x-t_0)}{\sqrt{2}\sigma} + \frac{\sigma}{\sqrt{2}\tau}\right]$ is differentiated. Computation of the resulting terms yields the analytical formula \eqref{Analyticfitfunc}.

\begin{figure}[htbp]
\includegraphics[width=.9\textwidth]{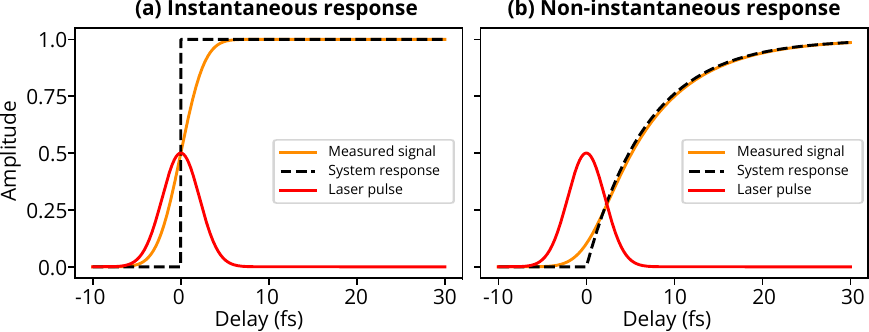}
\centering
\caption{\textbf{Non-instantaneous response} Both plots show the system's impulse response $h(t)$ (dotted black line), a Gaussian excitation pulse (red solid line) and the resulting measured signal (yellow solid line). \textbf{a,} When  $\tau=0$, the measured signal is an error function,  looking similar to the TDDFT+U calculations (Fig. 4 of the main text).   A slow characteristic time (here, $\tau=$\qty{7}{\femto\s}) yields a non-symmetric signal similar to the experimental data (Fig. 4 of the main text).} 
\label{fig:response}
\end{figure}

To better visualize the effect of the non-instantaneous response of the system, an illustration is shown on Figure \ref{fig:response} for two different cases. Figure \ref{fig:response}\textbf{a} shows the case of a system with an instantaneous response by setting the characteristic time  $\tau$ to $0$  while keeping the same Gaussian pump pulse width. As can be seen in eq \eqref{Analyticfitfunc}
when $\tau$ goes to $0$, the signal converges to a simple error function $ \frac{1}{2}   \text{erfc}\left(-\frac{t - t_0}{\sigma\sqrt{2}}\right) =  \frac{1}{2} \left[  1+ \text{erf}\left(\frac{t - t_0}{\sigma\sqrt{2}}\right) \right]$. This case is close the result of TDDFT+U calculations where the system responds instantaneously to the driving laser (Figure 4 of the main text). 

Figure \ref{fig:response}\textbf{b} instead shows the case of an idealized non-instantaneous response, $h(t,t_0,\tau)$, with $\tau=$ \qty{7}{\femto\s} and $t_0=0$; the Gaussian pulse of width $\sigma=$ \qty{5}{\femto\s} used for excitation, and the resulting signal that would be obtained. This yellow curve is similar to the one measured in the experiment and illustrates that the obtained signal is not symmetric around $t=0$. The differences between non-instantaneous and instantaneous reponses (panel (a) and (b), respectively) is very often overlooked in experiments. However, in the present case with $\sigma$ on the order of magnitude as $\tau$, it is important to take into account to get physically meaningful results. We expect this approach, initially proposed in Ref \cite{deRoulet2024}, to be relevant for most attosecond transient absorption and reflectivity experiments.

\begin{figure}[htbp]
\includegraphics[width=.9\textwidth]{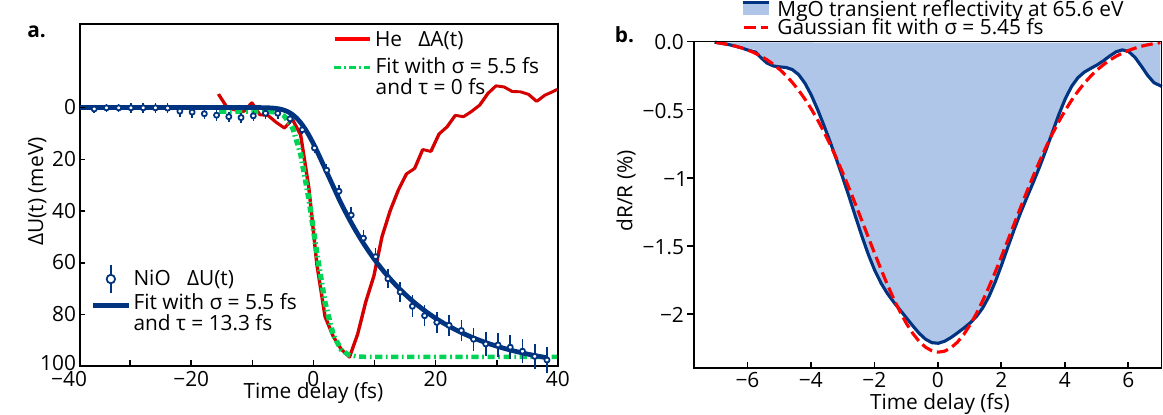}
\centering
\caption{\textbf{Instrumental time resolution} \textbf{a,} A lineout of the He \textit{2s2p} transient absorption (red curve, arbitrary units), is overlaid with $\Delta U(t)$ from NiO (same data as Fig. S4 of the main manuscript). A fit of the He signal (green dash-dotted curve) shows a $\sigma=$\qty{5.5}{\femto\s}.
\textbf{b,} Lineout of the MgO transient reflectivity signal at 65.6~eV, filtered to remove most of the $2\omega$ oscillations (shaded blue area) and gaussian fit (dashed red line), showing that the time resolution is $\sigma=$\qty{5.5}{\femto\s} as well.}
\label{fig:resolution}
\end{figure}

In order to fit the data, the values of $t_0$ and $\sigma$ must be well characterized. We do this in two separate ways for consistency. The first method relies on a transient absorption scan performed in helium, right before the NiO measurement. As shown in Fig.~\ref{fig:resolution}a, we take a lineout at the energy of the $2s2p$ resonance, and fit one side of the signal with an instantaneous response (the other side is linked to the lifetime of the core-excited state). This yields $\sigma$ = \qty{5.5}{\femto\s}. The second method relies on the MgO measurement. It is known that away from the excitonic resonances, the signal lasts only during the time overlap of the pump and probe pulse, therefore it reflects the instrumental time resolution \cite{geneaux_attosecond_2020}. Therefore, we filter out the fast oscillations and study the envelope of the signal at an arbitrary XUV energy. Fig.~\ref{fig:resolution}b shows the signal together with a gaussian fit, unveiling a resolution of $\sigma=$ \qty{5.45}{\femto\s}. Repeating this procedure for other energies yields consistent values of $\sigma=$ \qtyrange{5.1}{5.5}{\femto\s}. Taken together, both methods guarantee that $\sigma =$~\qty{5.5 \pm 1.0}{\femto s} is a safe and reliable estimate for the instrumental resolution.

\noindent $\tau$ then remains the only undetermined parameter. Its value is then obtained by fitting $\Delta U(t)$ with Equation~\ref{Analyticfitfunc}. The uncertainty on $\sigma$ is propagated by a Monte-Carlo procedure, yielding a \qty{\pm 1.0}{\femto \s} error on $\tau$. The final value for $\tau$ is thus \qty{13.3 \pm 1.0}{\femto \s}.

\section{Behavior at longer timescale}
\begin{figure}[hbp]
\includegraphics[width=.8\columnwidth]{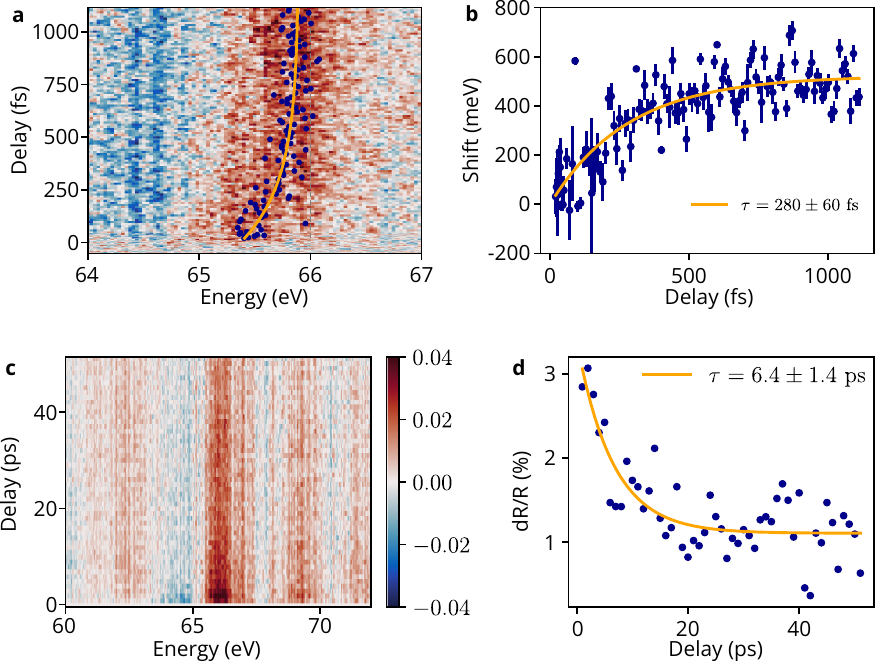}
\centering
\caption{\textbf{Long time behavior} \textbf{a,} The transient reflectivity measured up to \qty{1}{\pico\s}. The central position of the lowest energy feature is shown in blue along with a filtered curve to highlight the trend.   \textbf{b,}
Displacement of the first peak as a function of time fitted by an exponential decay curve yielding a characteristic time of \qty{280\pm60}{\femto\s}. \textbf{c,} The transient reflectivity measured up to \qty{50}{\pico\s}. \textbf{d,}~Amplitude of the transient reflectivity signal as a function of time, also fitted by an exponential decay curve with a characteristic time of \qty{6.4\pm1.4}{\pico\s}. The signal's persistence at a non-zero value indicates the presence of a metastable, long-lived state. }
\label{fig:S5}
\end{figure}

NiO presents a signal that lasts for a much longer time than its initial electronic response. We measured this response up to our maximal range of 50~ps, limited by our motorized delay stage range. After the initial rise of the redshift, which is associated with an electronic response time of \qty{13.3\pm1.0}{\femto\s}, NiO shows a different signal that appears as a blue shift of the transient reflectivity. It is shown on Figure \ref{fig:S5}a. 

\clearpage
This signal is of a different nature from the rigid shift of the static reflectivity curve discussed in the main manuscript. Here we observe a blue shift of the \textit{transient} signal itself, which corresponds to a more complex evolution of the material's reflectivity. To account for this shift, we introduce the displacement $D$ to track the center of the transient reflectivity peak. The results are shown on Figure \ref{fig:S5}b. We fit the resulting curve with an exponential decay function $Ae^{-t/\tau}+B$, yielding a characteristic decay time of $\tau_1$ = \qty{280 \pm 60}{\femto \s}. This timescale is consistent with the value measured using ultrafast electronic diffraction $\tau_1 \approx$ \qty{310 \pm 80}{\femto\s} reported in Ref \cite{Windsor2021}, indicating that this signal originates from the crystal lattice deformation. At longer timescale, the amplitude of the signal decays as shown on Figure \ref{fig:S5}b and d. The exponential decay fit yields a characteristic time of $\tau_2$ = \qty{6\pm 0.7}{\pico\s}, eventually stabilizing at a non-zero value. The reported value given by Ref \cite{Windsor2021} is $\tau_2$ = \qty{4.1 \pm 1.3}{\pico\s} which is compatible with our value, given the large uncertainties. This signal persists well beyond the experimental range of  $50$~ps. As no observable decay was detected, the signal may extend into the nanosecond regime. These results are identical to the one found with electronic diffraction \cite{Windsor2021}, allowing us to conclude that we observe the effect of the lattice deformation. \\

\begin{figure}[htbp]
\includegraphics[width=\columnwidth]{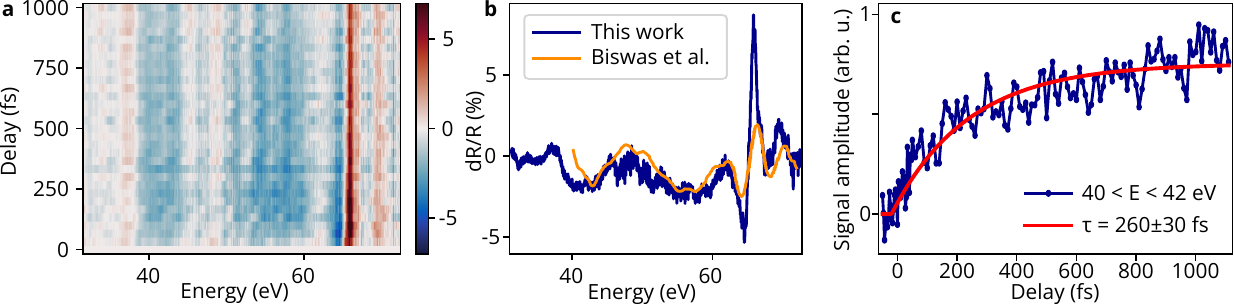}
\centering
\caption{\textbf{Low energy signal} \textbf{a,} The transient reflectivity measured up to \qty{1}{\pico\s} and shown across a wider energy range. \textbf{b,} Lineout of the transient reflectivity at positive delays, overlapped with the data from Ref.~\cite{Biswas2018}. 
\textbf{c,}~Evolution of the signal integrated between 40 and 42~eV, showing a \qty{260 \pm 30}{\femto \s} timescale.}
\label{fig:S5_Bis}
\end{figure}

Finally, we also captured the transient reflectivity signal at lower photon energies located below the Ni edge. Figure \ref{fig:S5_Bis}a shows the transient reflectivity, on which weaker features are visible on a wide energy range. Their shape matches exactly the ones measured by Biswas and coworkers \cite{Biswas2018} (Fig. \ref{fig:S5_Bis}b) who directly excited the charge-transfer gap of NiO. They were explained as absorption from oxygen levels, triggered by the formation of a localized excited state. The evolution of this low-energy signal (\ref{fig:S5_Bis}c) reveals a \qty{260 \pm 30}{\femto \s} timescale, very similar to the structural timescale mentioned above. Taken together, these concomitant electronic and structural dynamics show that the excited state of NiO at picosecond timescales is consistent with the formation of a polaron.

\clearpage
\section{Transient reflectivity at different excitation intensities}
\begin{figure}[h]
\includegraphics[width=1.0\columnwidth]{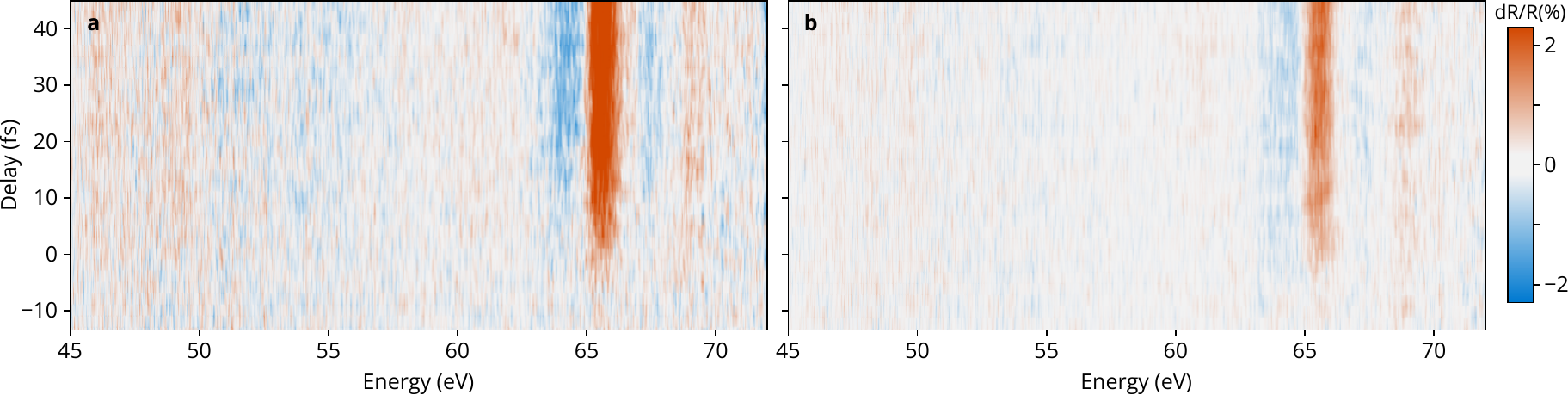}
\caption{Transient reflectivity taken at (\textbf{a}) \qty{6}{\tera\watt\per\square\cm} and (\textbf{b}) \qty{4}{\tera\watt\per\square\cm}.}
\label{fig:comparison}
\end{figure}

Figure \ref{fig:comparison} presents the transient reflectivity of NiO at \qty{6}{\tera\watt\per\square\cm} (data used in the main manuscript) and at \qty{4}{\tera\watt\per\square\cm} on the same color scale. The shape and the dynamics of the features appear extremely similar, with a smaller magnitude of $dR/R$ (about 0.55 of the high fluence data). Performing the same analysis in terms of $U$ change for the \qty{4}{\tera\watt\per\square\cm} data yields a final $\Delta U$ of \qty{50\pm10}{\milli\eV}. However, the lower magnitude of the signal makes the extraction of a quantitative timescale more challenging in this case.

\section{Comparison of static reflectivity obtained by various methods}

Here we compare the experimentally measured static reflectivity, together with the ones calculated by either multiplet calculations or TDDFT+U theory (see Methods of main manuscript for complete details). 
\begin{wrapfigure}{r}{0.6\textwidth}
    \vspace{0pt} % adjust vertical placement
    \includegraphics[width=0.55\textwidth]{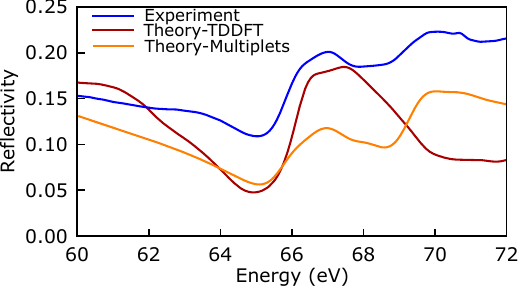}
    \caption{Static reflectivity from experiment (blue) compared to the multiplet theory (orange) and to TDDFT theory (brown).}
    \label{fig:S6}
\end{wrapfigure}
Because its parameters were adjusted to fit the experiment, the multiplet is able to reproduce very well the experiment, up to an offset (set to match literature data \cite{henke1993x}, see Section \ref{sec:KK}). On the contrary, TDDFT+U is entirely \textit{ab initio}, but is still able to reproduce the first peak very well, which is where the transient signal occurs.

\clearpage
\section{Carrier envelope phase dependence of the signal in TDDFT+U simulations}
\begin{figure}[h]
\includegraphics[width=0.6\columnwidth]{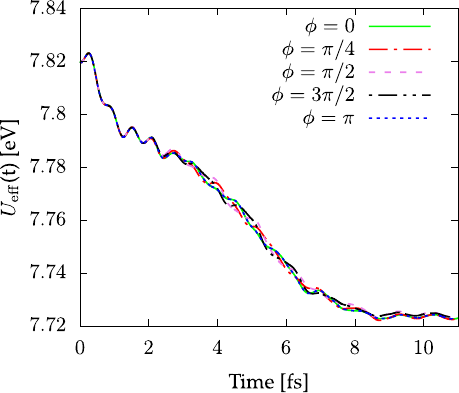}
\caption{Time-dependent $U_{\rm eff}(t)$ calculated by TDDFT+U for several carrier-envelope phases of the pump pulse.}
\label{fig:cep}
\end{figure}

Both experiment and theory presented in the main manuscript do not show signs of electric-field dependence of the NiO signal. In order to further strengthen this observation, we investigated the impact of the carrier envelope phase (CEP) in the TDDFT+U calculations. Here the calculations are performed with the pump-only and the effective Hubbard parameter $U_{\rm eff}(t)$ is obtained as a function of time, as shown in Fig.\ref{fig:cep}.\\ 

Overall, the CEP is found to have a negligible effect on $U_{\rm eff}(t)$. This confirms that the U renormalization is intensity-driven, and not field-driven, and also guarantees that the absence of CEP stabilization in the experiment does not have consequences on the measured signal.

\clearpage

\section{Multiplet calculations}
Here we detail how multiplet calculations were performed. Following the strategy outlined in the main article, the reduction of the Hubbard $U$ is accompanied by a decrease in charge-transfer energy $\Delta E_{CT}$ equal to $\Delta U/2$. Since the initial values considered for the unperturbed case are $U = \SI{7.3}{\electronvolt}$ and $E_{CT} = \SI{4.7}{\electronvolt}$, the above constraint results in: $U=7.3-2(4.7-E_{CT})$, the values being given in \si{\electronvolt}. 

\begin{figure}[htbp]
\includegraphics[width=\textwidth]{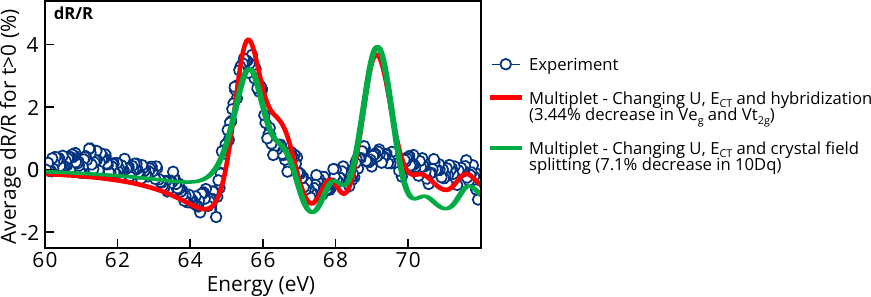}
\centering
 \caption{Comparison of the experimental results with multiplet calculations taking into account the change of hybridization vs. crystal field splitting}
\label{fig:change_10Dq}
\end{figure}

As indicated in the main manuscript, changes of $U$ and $E_{CT}$ are not sufficient to reproduce the experimental spectrum. Thus we considered modifying either the hybridization parameters, V$_{e_g}$ and $V_{t_{2g}}$, or the crystal field splitting, $10Dq$. Figure \ref{fig:change_10Dq} shows the best results by combining a change in $U$ with either a reduction of $10Dq$ (green line), or a reduction of $Ve_g/t_{2g}$ (red line). Both parameterizations reproduce the experimental spectra comparably well. The observable therefore does not provide sufficient sensitivity to uniquely determine which of the two parameters is responsible for the observed spectral evolution. 

The choice between these parameterizations must therefore rely on physical considerations. A change in $10Dq$ primarily reflects a modification of the local crystal field, which is strongly influenced by the Ni–O bond length and is therefore expected to occur on lattice timescales. By contrast, the hybridization is directly impacted by light-induced changes in electronic populations. Since we are interested in fitting the the spectrum in the first few femtoseconds following excitation, we consider varying only the hybridization to be the more physically appropriate description. We keep $10Dq$ at its fixed value of \SI{0.56}{\electronvolt}.

The hybridization parameters V$_{e_g}$ and $V_{t_{2g}}$ were considered related through $V_{e_g}=1.7\ V_{t_{2g}}$ \cite{haverkort2012multiplet}. We further considered a fixed reduction of the Slater integrals. With these considerations, the number of variable parameters for the multiplet calculations can be reduced to only two: the charge-transfer energy $E_{CT}$ and the hybridization $V_{t_{2g}}$. We considered 25 hybridization values equally distributed over the interval $\SI{1.96}{\electronvolt} <V_{eg}< \SI{2.06}{\electronvolt}$. For each $V_{eg}$ value, we extracted the best shift for 150 random $E_{CT}$ values considered from $E_{CT}^{min} = \SI{4.5}{\electronvolt}$ to $E_{CT}^{max} = \SI{4.7}{\electronvolt}$. Then, we
evaluated which of these combinations delivered the most reliable red shift. The redshift is determined in the same way as in Ref \ref{redshift_analysis}. We found that the values of the shift were distributed from around \SI{-0.077}{\electronvolt} to \SI{0}{\electronvolt}.
\begin{figure}[h!]
    \includegraphics[width=96mm]{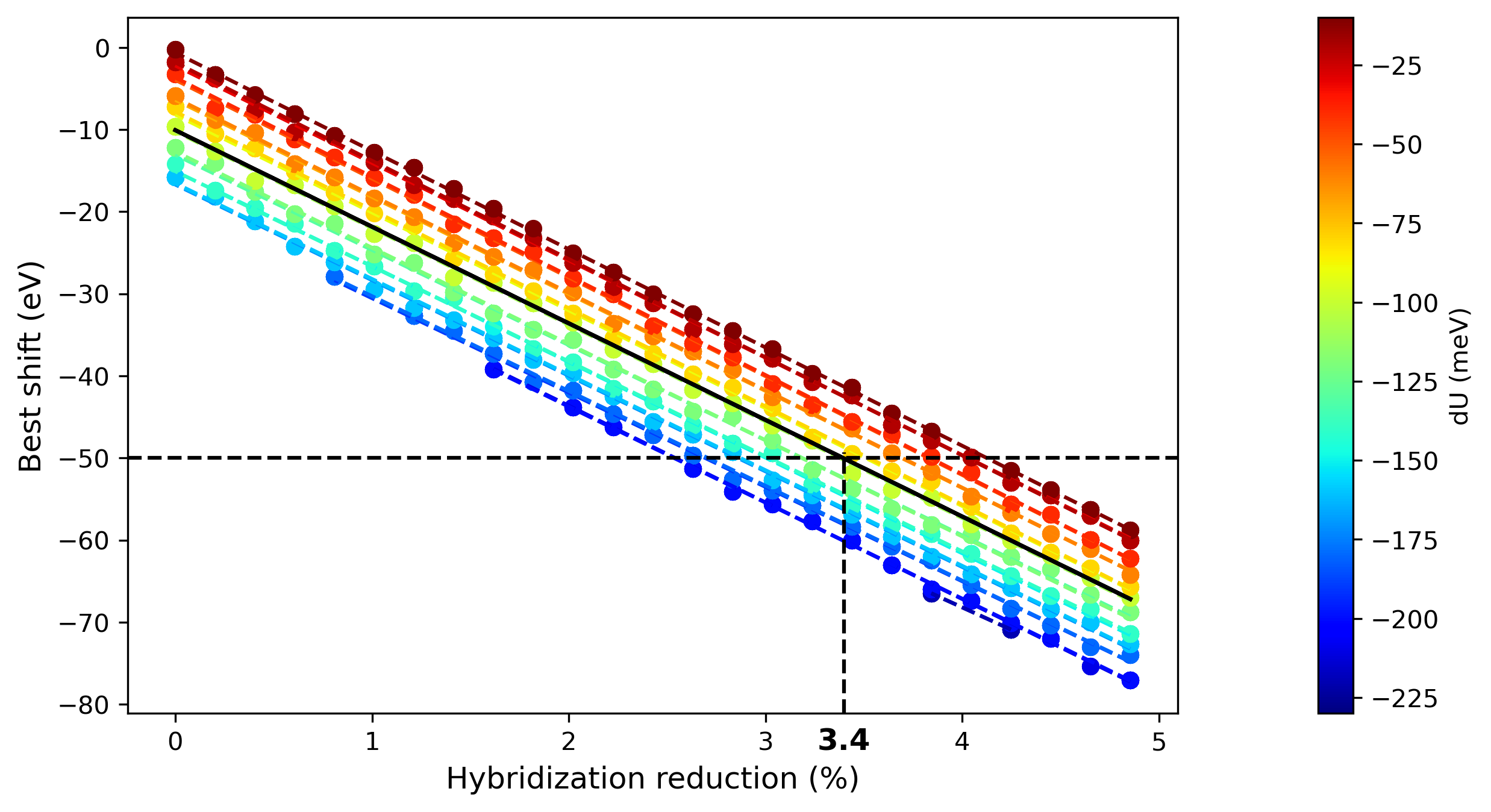}
    \centering
    \caption{For each value of dU (reduction of on-site Coulomb repulsion), we can draw a linear trend  for best shift vs hybridization percentage reduction. The best shifts are grouped by \SI{2}{\milli\electronvolt}. We can see that for the shift of \SI{-50}{\milli\electronvolt} and the dU = \SI{-100}{\milli\electronvolt}, we obtain an hybridization reduction of 3.4 \%.}
    \label{fig:hyb_vs_bestshift}
\end{figure}
In order to ease the comparison, in Figure~\ref{fig:hyb_vs_bestshift}, we show the obtained shifts as a function of hybridization reduction for each d$U$. The hybridization reduction is represented as a percent change of its initial value $V_{e_g} =\SI{2.06}{\electronvolt}$. A general trend can be observed: for all d$U$ values, the redshift amount increases linearly with the hybridization. Moreover, for a given redshift, there is 
an ensemble of solutions (d$U$, hybridization reduction). Considering the experimentally measured redshift (\SI{-50}{\milli\electronvolt}) and the d$U$ obtained by TDDFT+U (\SI{-100}{\milli\electronvolt}), shown as dashed lines in \ref{fig:hyb_vs_bestshift}, our calculations yield an hybridization reduction of \SI{3.4}{\percent}.

\clearpage
\bibliography{references}